\documentclass[pre,twocolumn,floatfix,showpacs,superscriptaddress,aps]{revtex4}

\usepackage{amsmath,graphicx}
\usepackage{color}

\begin{document}

\title{Scaling and synchronization in a ring of diffusively coupled nonlinear oscillators}

\author{D.~V.~Senthilkumar}
\affiliation{ Centre for Dynamics of Complex Systems, University of Potsdam, 14469 Potsdam Germany}%
\affiliation{Potsdam Institute for Climate Impact Research, 14473 Potsdam Germany}%
\author{P. Muruganandam}%
\affiliation{Department of Physics, Bharathidasan University, Tiruchirapalli - 620 024, India}
\author{M.~Lakshmanan}%
\affiliation{Centre for Nonlinear Dynamics, Bharathidasan University, Tiruchirapalli - 620 024, India}%
\author{J.~Kurths}%
\affiliation{Potsdam  Institute for Climate Impact Research, 14473 Potsdam Germany}%
\affiliation{Institute for Physics, Humboldt University, 12489 Berlin, Germany}%


\begin{abstract}

Chaos synchronization in a ring of diffusively coupled nonlinear oscillators 
driven by an external identical oscillator is studied. Based on numerical 
simulations we show that by introducing additional couplings at $(mN_c+1)$-th 
oscillators in the ring, where $m$ is an integer  and $N_c$ is the maximum 
number of synchronized oscillators in the ring with a single coupling, the 
maximum number of oscillators that can be synchronized can be increased 
considerably beyond the limit restricted by size instability. We also demonstrate that there exists an exponential relation between the number of oscillators 
that can support stable synchronization in the ring with the external drive 
and the critical coupling strength $\varepsilon_c$ with a scaling exponent 
$\gamma$. The critical coupling strength is calculated by numerically 
estimating the synchronization error and is also confirmed from the conditional 
Lyapunov exponents (CLEs) of the coupled systems. We find that the same scaling 
relation exists for $m$ couplings between the drive and the ring. Further, we 
have examined the robustness of the synchronous states against Gaussian 
white noise and found that the synchronization error exhibits a power-law 
decay as a function of the noise intensity indicating the existence of both noise-enhanced 
and noise-induced synchronizations depending on the value of the coupling strength
$\varepsilon$. In addition, 
we have found that $\varepsilon_c$ shows an exponential decay as a function 
of the number of additional couplings. These results are demonstrated using 
the paradigmatic models of R\"ossler and Lorenz oscillators.

\end{abstract}

\maketitle

\section{Introduction}

Chaos synchronization has been receiving a great deal of interest for
more than three decades \cite{fuji:83:01,fuji:83:02, pecora:90:01,
Pikovsky_etal:book:01, lakshman_murali:book:96}. In particular, chaos
synchronization in arrays  of coupled nonlinear dynamical systems
has been extensively investigated over the years in view of its
diverse applications in spatially extended systems, neural process,
networks, etc. \cite{bohr:89:01, heagy:94:01, kocarev:94:01,
gerson_anand:03-01,arenas:08:01,osipov:07:01}. Linear arrays with periodic boundary condition
(ring geometry) have been used widely in modeling
physiological, biochemical and biological phenomena
\cite{osipov:07:01,matias:97-01, matias:98-01}. For example, morphogenesis in
biological context \cite{turing:52-01} and transitions between different
animal gaits have been explained by considering a model composed of
a ring of coupled oscillators \cite{collins:94-01}. An important
application of the ring geometry is that the resulting spatiotemporal
patterns in the ensemble of coupled oscillators can be analyzed
through symmetry arguments \cite{collins:94-01, matias:97-01}.
Recently several interesting dynamical properties/collective behaviors including
amplitude death and chimera states have been identified
in such a ring type configuration~\cite{osipov:07:01,matias:97-01, matias:98-01,reddy_sen:04-01,abrams_strogatz:04-01,deng_huang:02-01, 
wang_huang:05-01,sande_soriano:08-01,
jiang_etal:03-01, lorenzo:96-01,harmov:06-01,brandt:06-01}.

Some of the recent studies have considered synchronization dynamics
in both ring and linear arrays coupled together in order to 
understand the dynamics of basic units of networks \cite{osipov:07:01,
deng_huang:02-01, wang_huang:05-01,sande_soriano:08-01}. Recently,
interesting scaling behavior of correlation properties of 
interacting dynamical systems in such a configuration has been
demonstrated~\cite{sande_soriano:08-01}. However, most of the
studies have considered unidirectional coupling in both
the ring and linear arrays. Because of the diverse nature of interaction
in real world phenomena,  we have considered diffusively (nearest-neighbor)
coupled chaotic systems with ring  geometry driven by an external
identical drive in view of its widespread applications
in engineering, robotics, networks, and physiological and biological
systems~\cite{boccaletti:02-01, rosenblum:04-01}. 
For instance, cultured networks of heart cells are examples of biological structures with strong nearest-neighbor coupling \cite{soen:99-01, brandt:06-01}.

The phenomenon of size instability, where a critical size of the number 
of oscillators upto which a stable synchronous chaotic state exists, 
of a uniform synchronous state in arrays of coupled oscillators with 
both periodic and free-end boundary conditions have been widely studied
 ~\cite{pecora2, pecora:98:01, heagy:94:01, matias:98-01,
matias:97-01, bohr:89:01, restrepo:04:01, muruganandam:99:01, muru:thesis, 
palani_muru_lak:05:01}.
Increasing the number of oscillators beyond this limit leads to desynchronization 
and the occurrence of spatially incoherent behavior (eg., high-dimensional or spatiotemporal chaos). The stability of synchronous chaos in coupled
dynamical systems plays a crucial role in the study of pattern
formation, spatiotemporal chaos,  etc.~\cite{heagy:94:01,
muruganandam:99:01, restrepo:04:01, muru:thesis, rangarajan:02:01,
chen:03:01}. In this connection, using the paradigmatic models 
of R\"ossler and Lorenz oscillators we shall demonstrate in this paper that the maximal
number of oscillators in the ring geometry that can support stable
synchronous chaos can be increased by integer multiples of the original 
number of oscillators in the ring with additional couplings from the same drive oscillator.
Furthermore, it is also found that the critical coupling strength
and the number of oscillators which can be synchronized in the ring exhibit
an exponential relation with a scaling exponent and 
indeed the relation remains unaltered even on increasing
the number of couplings  between the drive and the ring.
In addition, the synchronization error displays a power law 
decay as a function of the noise intensity for a fixed value of
the coupling strength indicating the existence of noise 
enhanced/induced synchronization.
It is to be noted that small world networks can be generated by introducing
additional couplings between randomly selected nodes to create shortest
paths (links)  between distant nodes~\cite{arenas:08:01}. Further, recent
studies on synchronizability of networks have been employing pinning control
in which hubs in the networks are connected to the same drive node~\cite{arenas:08:01} as
in our present study.

In particular, we consider here the R\"ossler and Lorenz
oscillators in a ring geometry with diffusive coupling
between them  and driven by an external identical oscillator,
whose strength is proportional to a parameter $\varepsilon$ 
(see Eq.~(2) below). Based on
numerical simulations, we find that the critical coupling strength,
say $\varepsilon_c$, below which no synchronization exists $(\varepsilon<\varepsilon_c)$,
of the external drive increases exponentially
with a scaling exponent, $\gamma \in(0.3,0.5)$, as a function of
the number of oscillators in the array that supports a stable
synchronous state. Further we observe that the number of oscillators
which supports such a stable synchronous state can be increased in integer
multiples by introducing additional couplings at the $(mN_c+1)$-th oscillator
of the ring with same value of the coupling strength, where $m$ is the number of 
couplings and $N_c$ is the maximum number of
oscillators in the ring that can sustain stable synchronization with a 
single coupling. Interestingly, this exponential
relation is maintained while increasing the number of couplings, $m$, between
the array and the external drive. 
In addition, we have found that
$\varepsilon_c$ shows an exponential decay as a function of the
number of additional couplings between the drive and the response array
for a fixed number of oscillators in the array.
Further, we find that these results are robust against Gaussian
white noise of small intensity  and the synchronization error
exhibits a power law decay as a function of the noise intensity. These results also indicate the existence of 
both the phenomena of noise-enhanced 
and noise-induced synchronizations depending on the value of the 
coupling strength $\varepsilon$ beyond certain 
threshold values of the noise intensity.  It is to be
emphasized that all the numerical simulations throughout the
manuscript have been repeated with several initial conditions and 
the results are the averages of a large number of realizations. 

The structure of the paper is as follows. In Sec.~\ref{sec2}, we study
synchronization in rings of diffusively coupled R\"ossler and Lorenz
systems driven by external identical oscillators with a drive-response
configuration. We show that in both the cases the critical coupling strength
increases exponentially with the number of oscillators in the response
array with a scaling exponent. In addition, the systems exhibit a power law decay
of the synchronization error as a function of the noise intensity demonstrating
noise-enhanced and noise-induced synchronizations depending on the value of
the coupling strength.
In Sec.~\ref{sec3}, we demonstrate
that the size of the ring can be increased beyond the size instability limit 
by integer multiples of the
maximum number of synchronized oscillators ($N_c$) in the ring with a single coupling
for the same value of coupling strength.
We also show that the same scaling relation with a characteristic
exponent is valid for the case of arbitrary number of couplings also. Further, we examine
the effect of noise on the robustness of the synchronous state for a fixed value of
the noise intensity and as a function of the noise intensity for a fixed
value of the coupling strength in all the cases.
Finally, in Sec.~\ref{sec4}, we present a summary and conclusions.

\section{Chaos synchronization in diffusively coupled
oscillators driven by an external identical oscillator}
\label{sec2}

\begin{figure}[!ht]
%\centering\includegraphics[width=0.8\linewidth]{fig1}
\centering\includegraphics[width=0.5\linewidth]{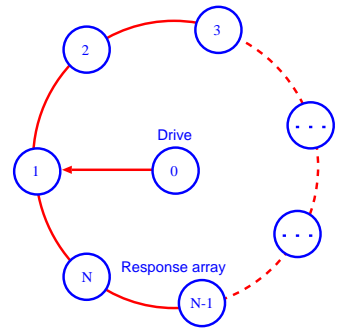}
\caption{(color online) Schematic diagram of the ring of diffusively coupled
oscillators driven by an external identical oscillator with
drive-response configuration.} \label{fig_arr}
\end{figure}

For the present study, we consider a coupling scheme with drive-response
configuration as shown in Fig.~\ref{fig_arr}, in which the diffusively
coupled circular array is driven by an external identical oscillator.
Here `$0$' denotes the external drive and `$1$', `$2$',
\ldots `$N$' denote the constituents of the response array with
nearest neighbor (diffusive) coupling. For simplicity, we assume that
all the oscillators in the drive-response configuration are identical.
In this section, we study chaos synchronization in the response array
when driven by the external drive. For this purpose, we have
considered R\"ossler and Lorenz oscillators coupled
according to Fig.~\ref{fig_arr}. The value of the diffusive coupling
constant is chosen such that all the oscillators in the array are in a
stable synchronous state. By varying the number of oscillators in the
response array from $N=2$ onwards, we calculate the critical value of the coupling constant
$\varepsilon_c$ at and above which the response array evolves in synchrony
with the external drive.

\subsection{Coupled R\"ossler system}
\label{secros}

First we analyse the coupled R\"ossler systems given by the dynamical equations
\begin{align}\label{eqn:ros}
\dot x_0 = &\, -(y_0+z_0), \notag \\
\dot y_0 = &\,  x_0+ay_0 \notag \\
\dot z_0 = &\,  b+z_0(x_0-c),
\end{align}
\begin{subequations}
\label{eqn:ros_arr0}
\begin{align}
\dot x_j = &\, -(y_j+z_j) \notag \\
           &\, + d(x_{j+1}+x_{j-1}-2x_j) + \delta_{1,j} \varepsilon (x_0 - x_j),\\
\dot y_j = &\, x_j+ay_j \\
\dot z_j = &\, b+z_j(x_j-c),\;\; j = 1,2,\cdots, N,
\end{align}
\end{subequations}
where $a=0.15$, $b=0.2$, $c=10$, $d=1$ and $\varepsilon$ is the coupling strength. 
Here the variable $x_0$, $y_0$ and
$z_0$ correspond to the drive system and $x_j$, $y_j$ and
$z_j$ $(j=1,2,\ldots,N)$ represent the diffusively coupled response
array. The first oscillator of the response array
(\ref{eqn:ros_arr0}) is driven by the drive (\ref{eqn:ros}).
$\delta_{i,j}$ is the Kronecker delta function given by
\[
\delta_{i,j} = \left\{ \begin{array}{ll} 1 & \text{ for } i = j \\
0 & \text{ otherwise}.
\end{array}\right.
\]
The isolated system (\ref{eqn:ros}) exhibits chaotic behavior
for the above choice of parameters with the Lyapunov exponents
$\lambda_1 \approx 0.1304> 0$, $\lambda_2 = 0$ and $\lambda_3 \approx
-14.1405$.

Next we shall study the dynamics of the drive-response configuration
(\ref{eqn:ros}) and (\ref{eqn:ros_arr0}). The value of the
diffusive coupling constant $d$ is chosen such that it supports
the maximum number of oscillators in the circular array to evolve
in synchronous fashion. In the absence of any external coupling
($\epsilon=0$), the array of diffusively coupled R\"ossler systems
(\ref{eqn:ros_arr0}) exhibits chaos synchronization in all the $N$
oscillators with $N \le N_{\mbox{max}}$ with~\cite{heagy:94:01}
\begin{align}
N_{\mbox{max}}=\frac{\pi}{\sin^{-1}\left(\sqrt{\lambda_{\mbox{max}}/4d}\right)},
\label{nmax0}
\end{align}
where $\lambda_{\mbox{max}}$ is the maximal Lyapunov exponent.
One can easily check that $N_{\mbox{max}} = 17$ for the above choice of parameters.

As soon as the external drive is switched on ($\epsilon \ne 0$),
the stable synchronous state of the diffusively coupled
oscillators gets destroyed for small values of $\epsilon$. The
number of oscillators in the response array (\ref{eqn:ros_arr0})
which retains synchronization with the drive (\ref{eqn:ros})
depends on the coupling strength $\epsilon$.  To estimate the
quality of synchronization, we define a measure, namely the
synchronization error, as the Euclidean norm
\begin{align}
\eta(t) & = \frac{1}{N}\left\{\sum_{j=1}^{N}\left[(x_0-x_j)^2+(y_0-y_j)^2
+(z_0-z_j)^2\right]\right\}^{\displaystyle\frac{1}{2}},\label{eta0}
\end{align}
In order to get a perfect synchronization in the drive-response
configuration, (\ref{eqn:ros}) and (\ref{eqn:ros_arr0}), we require $\eta
\to 0$ as $t\to\infty$. We remark here that all the simulations
in the manuscript are performed for an average of over $100$ random initial
conditions. By examining $\eta$, we extract the
critical coupling strength $\varepsilon_c$ corresponding to the
number of oscillators, $N$, in the response array, whose dynamics
are entrained with that of the drive in phase space. For
example, for $N=2$, we find the critical coupling strength as
$\varepsilon_c = 0.47$. The $\varepsilon_c$ values are given in
Table~\ref{table1} for different values of $N$. However, for
$N \ge 6$, the system gets destabilized or becomes completely
unstable, for any value of $\varepsilon\ne 0$.
\begin{table}[!ht]
\begin{center}
\caption{Critical coupling strength $\varepsilon_c$}
\label{table1}
\begin{tabular}{rr|rr}
\hline \multicolumn{2}{c|}{R\"ossler system} &
\multicolumn{2}{c}{Lorenz
system} \\
$N$ & $\varepsilon_c$ & $N$ & $\varepsilon_c$ \\
\hline
$2$ & $0.447$ & $2$ &    $12.528$ \\
$3$ & $0.721$ & $3$ &    $20.288$ \\
$4$ & $1.134$ & $4$ &    $31.334$ \\
$5$ & $1.882$ & $5$ &    $51.486$ \\
\hline
\end{tabular}
\end{center}
\end{table}

By careful examination of the numerical values, we
realize that there is an exponential relation between the
number of oscillators ($N$) in the response array that supports
synchronization with the external drive and the critical coupling
strength, $\epsilon_c$. By fitting the numerical values, one can
easily establish the relation connecting $\epsilon_c$ and $N$ as
\begin{align}
\varepsilon_c = \varepsilon_0 \exp (\gamma N), \label{eqn:eps_c}
\end{align}
with the proportionality constant $\varepsilon_0 = 0.1665$ and the scaling exponent 
$\gamma = 0.4842$.
\begin{figure}[!ht]
\centering\includegraphics[width=0.99\linewidth,clip]{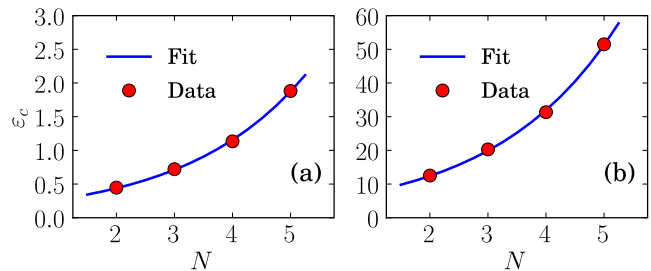}
\caption{(color online) Variation of the critical coupling
strength, $\varepsilon_c$, as a function of the number of
oscillators,
$N$, for (a) R\"ossler
system ($\varepsilon_0 =  0.1665$ and $\gamma = 0.4842$) and (b)
Lorenz system ($\varepsilon_0 = 4.8158$ and $\gamma = 0.4731$).
Filled circles correspond to the numerical data and solid lines
are the plot of $\varepsilon_c$ using relation (\ref{eqn:eps_c}).}
\label{fig:eps_c}
\end{figure}
The critical coupling strength as a function of the number of
oscillators is shown in Fig.~\ref{fig:eps_c}(a), where the filled circles
correspond to numerical data and the solid line is the fitted
data using the relation (\ref{eqn:eps_c}).
\begin{figure}[!ht]
%\centering\includegraphics[width=0.99\linewidth]{ros_lya}
\centering\includegraphics[width=\linewidth,clip]{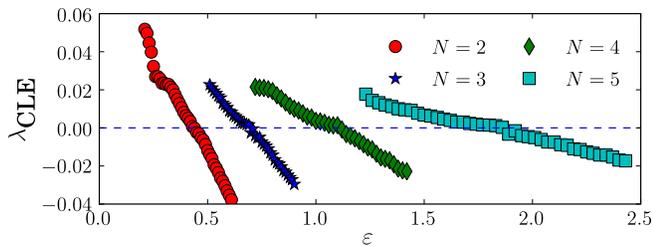}
\caption{(color online) Conditional Lyapunov exponents of the coupled R\"ossler system (\ref{eqn:ros_arr0}) for
different values of the number of oscillators ($N$) in the array as a function of the
coupling strength.} \label{ros_cle}
\end{figure}
The above result is also confirmed by calculating the conditional Lyapunov
exponents (CLEs) of the response array by increasing the number of the coupled
oscillators $N$. Here by CLE we mean the largest Lyapunov exponent of the response array, which of course should be less than zero in order to have a synchronous state. Fig.~\ref{ros_cle} shows the variation of CLEs as a function
of $\varepsilon$ for the various numbers of oscillators in the
response array.  From Fig.~\ref{ros_cle}, one can identify that the value of
$\varepsilon_c$ for each $N$, where a transition of the CLE from positive to
negative value occurs, agrees with the value of 
$\varepsilon_c$ given in Table~I. 

\subsection{Coupled Lorenz systems}

Next we consider the case of the Lorenz system described by the following
set of equations~\cite{sparrow:1982,lorenz:1963} as the drive,
\begin{align}
\dot x_0 = &\,  \sigma (y_0 - x_0 ), \notag \\
\dot y_0 = &\, r x_0-y_0-x_0 z_0  \notag \\
\dot z_0  = &\,  -\beta z_0+x_0 y_0, \label{eqn:lor}
\end{align}
where $\sigma = 10$, $r = 23$ and $\beta = 1$. A ring of
diffusively coupled Lorenz systems for the response array can be
represented by the following set of coupled equations,
\begin{subequations}
\label{eqn:clor}
\begin{align}
\dot x_j = &\,  \sigma (y_j - x_j )+d(x_{j+1}+x_{j-1}-2x_j) \notag \\
       &\, \qquad\qquad\qquad+ \delta_{1,j} \varepsilon (x_0 - x_j),\\
\dot y_j = &\, r x_j-y_j-x_j z_j  \\
\dot z_j = &\, -\beta z_j+x_j y_j,\;\; j = 1,2,\cdots, N,
\end{align}
\end{subequations}
where $d$ is the diffusive coupling constant and $\varepsilon$ is
the coupling strength of the external drive. One may note that the
isolated system (\ref{eqn:lor}) exhibits chaotic behavior for the
above choice of parameters with the Lyapunov exponents $\lambda_1
= 0.6075 > 0$, $\lambda_3 = 0$ and $\lambda_3 = -17.9194$.

We study the variation of the critical coupling strength as a function
of the number of oscillators in the response array by numerically
examining the synchronization error $\eta$ in a similar fashion as
in the case of the coupled R\"ossler oscillators. For this purpose, we
fix the diffusive coupling constant as $d=30$. In this case, the response
array (\ref{eqn:clor}), in the absence of external drive
$(\varepsilon = 0)$, supports a maximum of $N=9$ oscillators in
the stable synchronous state as per Eq.~(\ref{nmax0}).

When the external coupling is switched on, the synchronization  in
the array (\ref{eqn:clor}) gets lost. However, by choosing the
number of oscillators, $N$, and the coupling strength,
$\varepsilon$, appropriately one can synchronize the array with
the external drive. The values of 
$\varepsilon_c$ for different numbers of oscillators in the response
\begin{figure}[!ht]
\centering\includegraphics[width=\linewidth,clip]{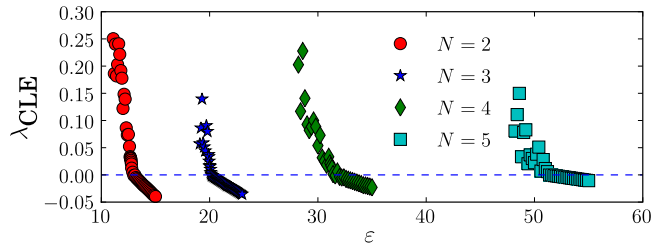}
\caption{(color online) Conditional Lyapunov exponents of the coupled Lorenz system (\ref{eqn:clor}) 
for different values of the number of oscillators in the array as a function 
of the coupling strength.} \label{lor_cle}
\end{figure}
array, $N$, are also given in Table~\ref{table1}.
Fig.~\ref{fig:eps_c}(b) depicts the plot of $N$ versus
$\varepsilon_c$. It is easy to see that, as in the case of the
coupled R\"ossler oscillators discussed above,
$\varepsilon_c$ again increases exponentially
as a function of $N$ according to the relation (\ref{eqn:eps_c})
with $\varepsilon_0 = 4.8158$ and  $\gamma = 0.4731$.

The above results have also been examined by calculating the conditional Lyapunov
exponents associated with (\ref{eqn:clor}) and it is confirmed that the largest 
CLE (cf. Fig.~\ref{lor_cle}) transits
from positive to negative value at the critical coupling strength
of a given $N$ as shown in Table~\ref{table1}.

So far, we have considered R\"ossler and Lorenz
oscillators with  a ring geometry and driven
by an external identical oscillator in a drive-response configuration.  Based on
numerical simulations, we have found that there is an exponential
relation connecting the critical value of the coupling strength $\varepsilon_c$
and the number of oscillators in the response array $N$ that evolve
in synchrony with the external drive with a scaling
exponent $\gamma \approx 0.48$ for  both the R\"ossler and Lorenz oscillators. 
Next, we will examine the effect of Gaussian white noise
on the stability of synchronization of the coupled oscillators.

\subsection{Effect of noise}

We have examined the robustness of the synchronous states of the coupled
oscillators by including Gaussian white noise to the first variable of all
the systems. Interestingly, we find that the results obtained remain unaltered
\begin{figure}[!ht]
\centering\includegraphics[width=0.99\linewidth,clip]{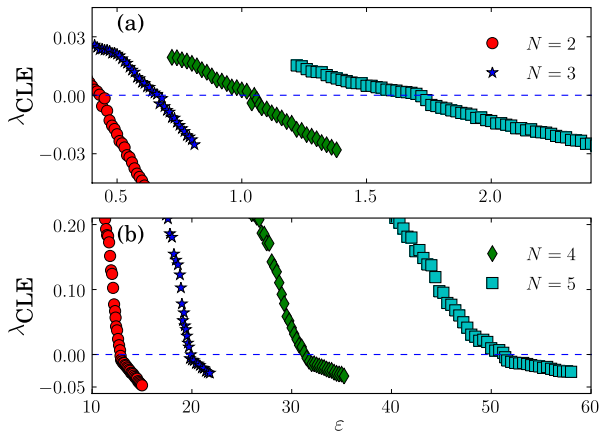}
\caption{(color online) Conditional Lyapunov exponents of the coupled (a) R\"ossler and, 
(b) Lorenz systems for different values of the number of oscillators in 
the array as a function of the coupling strength in the presence of 
Gaussian white noise with intensity $D_0=0.001$.} \label{single_nlya}
\end{figure}
for small values of the noise intensity and the synchronization error,
$\eta$, follows a power law decay as the intensity of the noise is increased
resulting in noise-enhanced and noise-induced synchronizations depending
on the value of the coupling strength. It is of interest to note that similar noise-enhanced
phase synchronization in two coupled noisy R\"ossler oscillators~\cite{osipov:07:01,zhou:02-01}
and noise-induced phase synchronization in two coupled R\"ossler and Lorenz 
oscillators~\cite{osipov:07:01,zhou:02-02} have been observed.

In particular, we have included the Gaussian white noise, $\sqrt{2aD_0}\xi(t)$ with
$a=0.01$	
to the $x$ variable of all the coupled systems including the drive after every
time step, where
$D_0$ is the noise intensity and $\xi(t)$ is the Gaussian white noise. We have
calculated the CLEs for both the R\"ossler and Lorenz oscillators as in the
previous section by including a small noise with noise intensity $D_0=0.001$, 
which are plotted in Figs.~\ref{single_nlya}(a) and ~\ref{single_nlya}(b), respectively. 
It is evident from this figure  that 
the critical values of $\varepsilon$ for the chosen value of
the noise intensity remain almost the same as in Figs.~\ref{ros_cle} and
~\ref{lor_cle} for the R\"ossler and Lorenz oscillators, respectively. Hence, the
exponential relation between the number of oscillators in the ring and their
corresponding $\varepsilon_c$ remains the same in the presence of
small noise also. 

\begin{figure}[!ht]
\centering\includegraphics[width=0.99\linewidth]{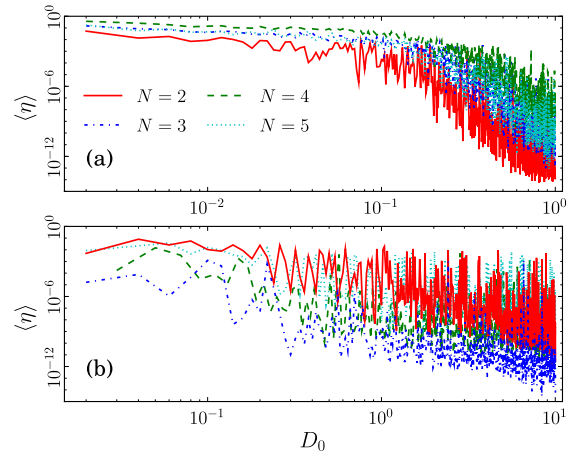}
\caption{(color online) Time averaged synchronization error $\langle \eta \rangle$ for different numbers of oscillators in the ring as a function of the noise intensity $D_0$  for a fixed value of $\varepsilon_c$ 
displaying a power law decay for (a) the R\"ossler oscillators in the range $D_0\in(0.01,1)$, and (b) the Lorenz oscillators in the range $D_0\in(0.01,10)$.}
\label{single_ni}
\end{figure}

Furthermore, we have calculated the synchronization error $\eta$ by increasing
the noise intensity $D_0$ at the threshold value of 
$\varepsilon$, namely $\varepsilon_c$, shown in Table~\ref{table1}.
The average synchronization error $\langle \eta \rangle$, where $\langle \cdot \rangle$ denotes
the time average over $500\, 000$ time steps, for different  values of $N$
in the ring as a function of noise intensity $D_0$ is shown
in Figs.~\ref{single_ni}(a) and ~\ref{single_ni}(b) for the R\"ossler and Lorenz systems, respectively. 
The average synchronization error for both the R\"ossler and Lorenz systems
for different numbers of oscillators in
the ring follows a power law decay as a function of the noise intensity $D_0$
beyond certain threshold values of $D_0$, i.e, $D_0 \ge 0.01$.

In our studies, we have fixed the value of the coupling strength $\varepsilon$ at the critical coupling strength $\varepsilon_c$ as shown in  Tables for different values of $N$. 
On increasing the noise intensity $D_0$, the synchronization error follows a power-law decay as a function of $D_0$ after certain threshold value of noise intensity. This naturally corresponds to a noise-enhanced synchronization and also confirms the robustness of the  synchronous state. Interestingly, we have also observed that by fixing $\varepsilon$ at a value less than that of the synchronization threshold $\varepsilon_c$, noise can induce synchronization between the ring and the drive oscillator. The synchronization error again shows a power-law decay as a function of the noise intensity (exactly similar to Figs.~\ref{single_ni} omitted here to avoid repetition), exhibiting noise-induced synchronization. Thus we have observed both the phenomena of noise-enhanced and noise-induced synchronizations depending upon the choice of $\varepsilon$.

\section{Overcoming size instability by introducing additional couplings}
\label{sec3}

\begin{figure}[!ht]
\centering\includegraphics[width=\linewidth]{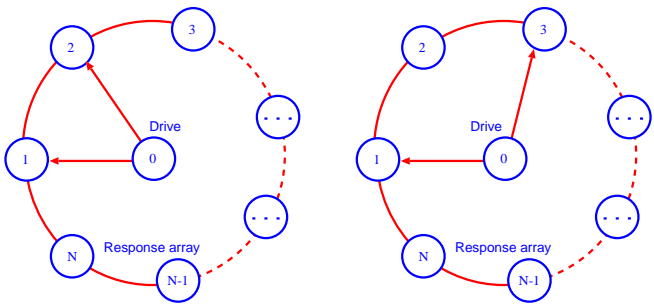}
\caption{(color online) Schematic diagram of the ring of diffusively coupled
oscillators driven by an external identical oscillator with additional 
coupling in (a) first neighbor and (b) second neighbor.} \label{fig_multi}
\end{figure}

In the above, we have pointed out that, when one considers an array of
diffusively coupled chaotic oscillators (with  a ring geometry) driven by
an external identical forcing in the drive-response configuration, the critical
coupling strength of the external drive required to synchronize the response array varies exponentially with the number of oscillators within the
synchronization regime. As a consequence of the exponential relation, the number
of oscillators in the response array that can be synchronized with the external
drive is limited to $4$ or $5$ because one requires a high coupling strength,
which results in desynchronization  above a certain threshold value of the
coupling strength due to size instability. However, we wish to point out
here that it is possible to increase the number of oscillators in the response
array, which are synchronized with the external drive by increasing the number of couplings. In Fig.~\ref{fig_multi} we show a schematic diagram 
for the realization of the drive-response configuration with more than one coupling. In this section we explore the possibility of increasing
the number of oscillators that evolve in synchrony with the drive and the
robustness of the scaling exponent $\gamma$ when one introduces more number of
couplings between the drive and the response array.

In order to study the effect of making more number of couplings
between the drive and the response array, we again consider the coupled
R\"ossler systems as discussed earlier in Sec.~\ref{secros}. The
drive system is assumed to follow the same set of equations (\ref{eqn:ros})
 as before. Then the
governing equations for the response array can be written as
\begin{subequations}
\label{eqn:ros_multi}
\begin{align}
\dot x_j  = &\, -(y_j+z_j) \notag \\
	     &\, +d(x_{j+1}+x_{j-1}-2x_j)  + p_j (x_0 - x_j),\\
\dot y_j  = &\,  x_j+ay_j, \\
\dot z_j  = &\,  b+z_j(x_j-c),\;\; j = 1,2,\cdots, N,
\end{align}
\end{subequations}
where
\begin{align}
p_j = \varepsilon\sum_{k=0}^{m-1}\delta_{j, kl+1},\;\;\;\;
\delta_{j,kl+1} = \left\{
\begin{array}{l}
1 \;\; \text{for} \;\; j=kl+1, \\
0 \;\; \text{otherwise}. \label{eqn:p}
\end{array}
\right.
\end{align}
and $\varepsilon$ corresponds to the coupling strength. Here $m$
denotes the number of couplings and $l$ represents the $l$-th
neighbor of the first oscillator in the response array. For example, $l=1$ for the
first neighbor, $l=2$ for the second neighbor and so on.

\subsection{Second coupling at the first neighbor}

First let us consider the case in which there is a second
coupling (of the same strength as the first one) between the external drive and 
the response array in the
immediate neighborhood of the first oscillator, that is,
$l=1$ in Eq.~(\ref{eqn:p}), which is already coupled to the drive [cf. Fig.~\ref{fig_multi}(a)].  
For simplicity, we consider the coupled R\"ossler
systems~(\ref{eqn:ros}) and (\ref{eqn:ros_multi}) with $N = 3$
with coupling at the first and second oscillators in the array.
Now, interestingly it is easy to see that the critical coupling strength
required to synchronize all the three oscillators in the array
gets reduced to $\varepsilon_c \approx 0.311$ from $\varepsilon_c\approx 0.407$ (see Table~\ref{table1}). Similarly, up to $N=6$ the second additional coupling 
helps to reduce  the critical values of $\varepsilon$ as tabulated in Table~\ref{table2} for the
\begin{table}[!ht]
\begin{center}
\caption{Critical coupling strength $\varepsilon_c$ for second coupling at the first neighbour}
\label{table2}
\begin{tabular}{rr|rr}
\hline 
\multicolumn{2}{c|}{R\"ossler system} 	& \multicolumn{2}{c}{Lorenz system} \\
$N$ & $\varepsilon_c$ & $N$ & $\varepsilon_c$ \\
\hline
$3$ & $0.311$ & $3$ &    $9.189$ \\
$4$ & $0.441$ & $4$ &    $12.895$ \\
$5$ & $0.647$ & $5$ &    $18.435$ \\
$6$ & $1.056$ & $6$ &    $28.297$ \\
\hline
\end{tabular}
\end{center}
\end{table}
coupled R\"ossler oscillators. The same analysis can be extended to the Lorenz oscillators (\ref{eqn:lor})-(\ref{eqn:clor}) as well. Again the results are tabulated in Table~\ref{table2}.  It may be noted that
for such a second additional coupling 
one additional oscillator, namely $N=6$, in the
ring can be synchronized compared to the case of a single coupling 
(see Table~\ref{table1}). Further,
the critical values of $\varepsilon$ in Table~\ref{table2} again show an exponential relation 
with the number of synchronized oscillators in the array as shown in
\begin{figure}[!ht]
\centering\includegraphics[width=0.99\linewidth,clip]{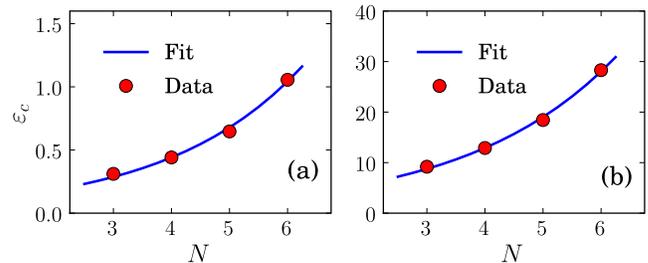}
\caption{(color online) The variation of the critical coupling
strength, $\varepsilon_c$, as a function of the  number of
oscillators,
$N$, for second additional coupling at the first neighbor of the first
coupling for (a) R\"ossler
system ($\varepsilon_0 = 0.0788$ and $\gamma = 0.43303$) and (b)
Lorenz system ($\varepsilon_0 = 2.7351$ and $\gamma = 0.3878$).
Filled circles corresponds to the numerical data and solid lines
are the plot of $\varepsilon_c$ using relation (\ref{eqn:eps_c}).}
\label{adcatfn_datafit}
\end{figure}
Figs.~\ref{adcatfn_datafit}(a)  and ~\ref{adcatfn_datafit}(b) for the
R\"ossler and Lorenz oscillators, respectively. Here the proportionality constant
$\varepsilon_0=0.0788$ and the scaling exponent $\gamma=0.4303$ for the R\"ossler oscillators
and $\varepsilon_0=2.7351$ and $\gamma=0.3878$ for the Lorenz oscillators.

\begin{figure}[!ht]
\centering\includegraphics[width=0.99\linewidth]{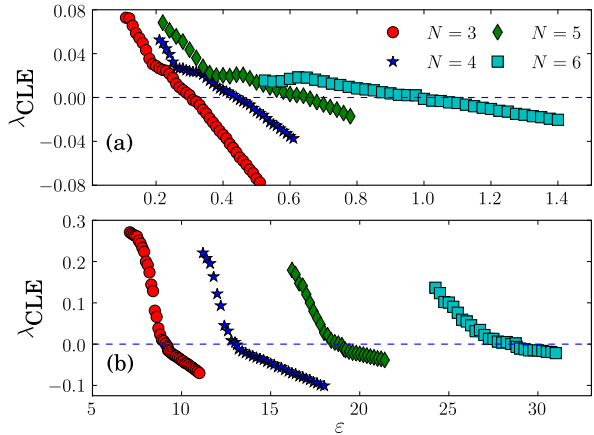}
\caption{(color online) Conditional Lyapunov exponents of the (a) R\"ossler and, (b) Lorenz systems  
for different values of the numbers of oscillators ($N$) in the array as a function 
of the coupling strength with the second additional coupling at the first neighbor
of the first coupling.} \label{adcatfn_lya}
\end{figure}

The conditional Lyapunov exponents of the coupled R\"ossler and  Lorenz 
oscillators for $N=3,4,5$ and $6$ oscillators in the ring as a function
of $\varepsilon$ is shown in Figs.~\ref{adcatfn_lya}(a) and~\ref{adcatfn_lya}(b),
respectively. It is to be noted that all the CLEs change their values
from positive to negative near the critical values of
$\varepsilon$ as in Table~\ref{table2}.
\begin{figure}[!ht]
\centering\includegraphics[width=0.99\linewidth]{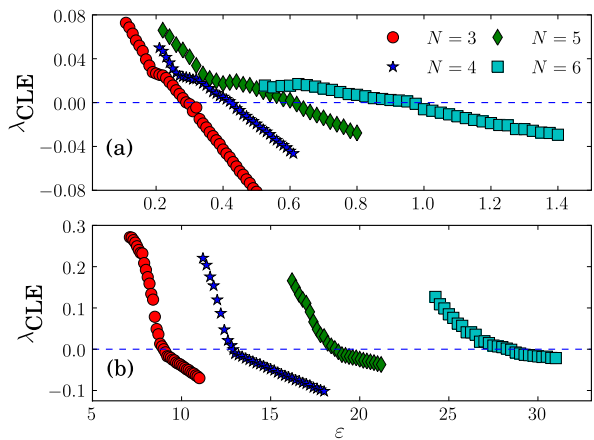}
\caption{(color online) Conditional Lyapunov exponents of the (a) R\"ossler and, (b) Lorenz systems  for different values of the number of oscillators ($N$) in the array as a function of the coupling strength with the second additional coupling at the first neighbor of the first coupling in the presence of Gaussian white noise with intensity $D_0=0.001$.} 
\label{adcatfn_nlya}
\end{figure}

We have also investigated the effect of noise in the present case as well. The CLEs of both the coupled R\"ossler and Lorenz oscillators for different $N$ values in the ring for the noise intensity $D_0=0.001$ is plotted in Figs.~\ref{adcatfn_nlya}(a) and \ref{adcatfn_nlya}(b) as a function of  $\varepsilon$.  Again the CLEs change their signs almost at the same critical values of $\varepsilon$ as the CLEs of the coupled oscillators without noise [Figs.~\ref{adcatfn_lya}(a) and \ref{adcatfn_lya}(b)] thereby preserving the same exponential relation between the critical values of  $\varepsilon$
and the number of oscillators synchronized in the ring. Further, 
\begin{figure}[!ht]
\centering\includegraphics[width=0.99\linewidth]{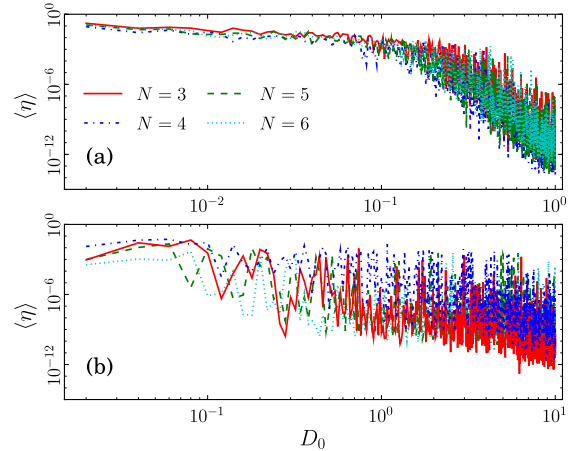}
%\centering\includegraphics[width=0.99\linewidth]{adcatfn_ni}
\caption{(color online) Time averaged synchronization error $\langle \eta \rangle$ for 
different number of oscillators in the ring as a function of noise intensity 
$D_0$  for fixed value of $\varepsilon_c$  displaying a power-law decay (a) the R\"ossler oscillators in the range $D_0\in(0.01,1)$, and the
Lorenz oscillators in the range $D_0\in(0.01,10)$ with the second additional
coupling at the first neighbor of the first coupling.} 
\label{adcatfn_ni}
\end{figure}
the synchronization error $\langle \eta \rangle$ follows a power law decay 
as a function of the noise intensity $D_0$ for both the R\"ossler and  Lorenz 
oscillators as shown in Figs.~\ref{adcatfn_ni}(a) and \ref{adcatfn_ni}(b) 
 beyond certain threshold values of $D_0$ for a fixed 
value of $\varepsilon_c$ given in Table~\ref{table2} illustrating noise-enhanced
synchronization. Further, a similar relation can be obtained between $\langle \eta \rangle$
and $D_0$  for values of $\varepsilon<\varepsilon_c$ confirming
the existence of noise-induced synchronizations.

\subsection{Second coupling at the second neighbor}

Next, by introducing a second coupling at the second neighbor (instead of the first one), $l=2$,
of the first coupling, we find that one can  synchronize two additional oscillators
in the ring compared to the number of synchronized oscillators with a 
single coupling. Likewise, introducing the second coupling at the third neighbor [cf. Fig.~\ref{fig_multi}(b)] of the first oscillator $(l = 3)$ in the
\begin{table}[!ht]
\begin{center}
\caption{Critical coupling strength $\varepsilon_c$ for second coupling at the second neighbour.}
\label{table3}
\begin{tabular}{rr|rr}
\hline \multicolumn{2}{c|}{R\"ossler system} &
\multicolumn{2}{c}{Lorenz
system} \\
$N$ & $\varepsilon_c$ & $N$ & $\varepsilon_c$ \\
\hline
$4$ & $0.405$ & $4$ &    $12.212$ \\
$5$ & $0.563$ & $5$ &    $16.231$ \\
$6$ & $0.814$ & $6$ &    $21.883$ \\
$7$ & $1.156$ & $7$ &    $31.987$ \\
\hline
\end{tabular}
\end{center}
\end{table}
array results in increasing the number of oscillators that are synchronized with the drive by $3$. Similarly, introducing the second coupling at the $N$-th neighbor will increase the synchronized oscillators in the ring by $N$. Thus from the Table~\ref{table1} for single coupling, we can realize that it is possible to synchronize up to $10$ oscillators (instead of $5$) with the introduction of a second coupling at reduced critical coupling strength.

\begin{figure}[!ht]
\centering\includegraphics[width=\linewidth,clip]{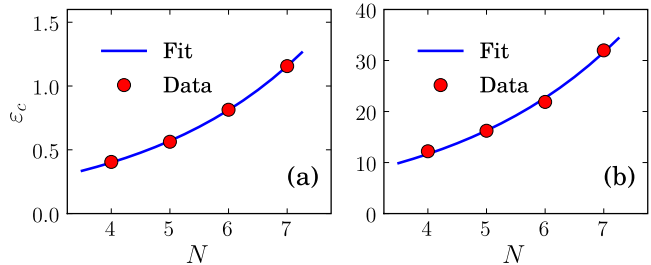}
\caption{(color online) The variation of the critical coupling
strength, $\varepsilon_c$, as a function of the number of
oscillators,
$N$, for second additional coupling at the second neighbor of the first
coupling for (a) R\"ossler
system ($\varepsilon_0 = 0.0972$ and $\gamma = 0.3536$) and (b)
Lorenz system ($\varepsilon_0 = 3.0908$ and $\gamma = 0.3321$).
Filled circles corresponds to the numerical data and solid lines
are the plot of $\varepsilon_c$ using relation (\ref{eqn:eps_c}).}
\label{adcatsn_datafit}
\end{figure}

\begin{figure}[!ht]
\centering\includegraphics[width=0.99\linewidth]{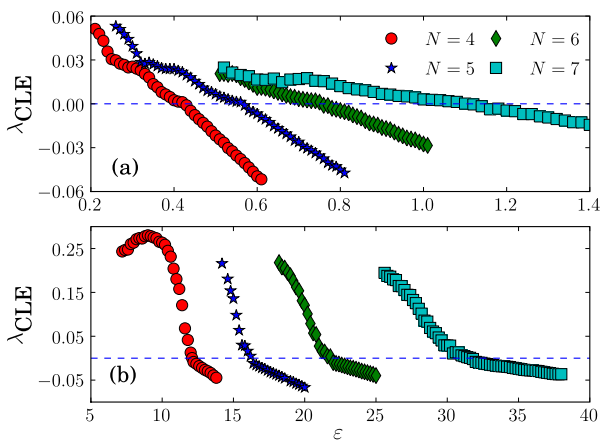}
\caption{(color online) Conditional Lyapunov exponents of the (a) R\"ossler and, (b) Lorenz systems for different values of the number of oscillators in the array as a function of the coupling strength with the second additional coupling at the second neighbor of the first coupling.} 
\label{adcatsn_lya}
\end{figure}

The critical values of the coupling strength $\varepsilon_c$
and their corresponding number of synchronized oscillators in the ring
for both the R\"ossler and  Lorenz oscillators  for the
second coupling at the second neighbor are tabulated in Table~\ref{table3},
which again displays an exponential relation as shown in Figs.~\ref{adcatsn_datafit}(a) and \ref{adcatsn_datafit}(b). 
The proportionality constant is estimated as
$\varepsilon_0=0.0972$ and the scaling exponent $\gamma= 0.3536$ for the R\"ossler oscillators
and $\varepsilon_0=3.0908$ and $\gamma=0.3321$ for the Lorenz oscillators.
These critical values of $\varepsilon$ are also confirmed by the transitions
in the value of the CLEs of the coupled R\"ossler and  Lorenz oscillators as shown in
Figs.~\ref{adcatsn_lya}(a) and \ref{adcatsn_lya}(b).
\begin{figure}[!ht]
\centering\includegraphics[width=0.99\linewidth]{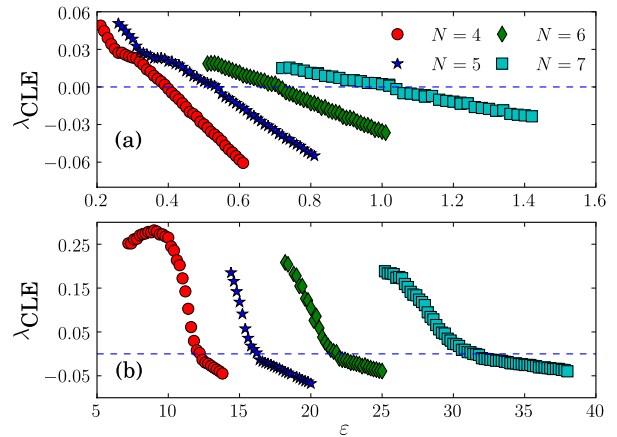}
\caption{(color online) Conditional Lyapunov exponents of the (a) R\"ossler and, (b) Lorenz systems for different values of the number of oscillators in the array as a function of the coupling strength with the second additional coupling at the second neighbor of the first coupling in the presence of the Gaussian white noise with
noise intensity $D_0=0.001$.} 
\label{adcatsn_nlya}
\end{figure}

The conditional Lyapunov exponents of both the oscillators as a function of
$\varepsilon$ with the noise intensity $D_0=0.001$ for $N=4,5,6$ and $7$ oscillators in the ring are shown in Figs.~\ref{adcatsn_nlya}(a) and \ref{adcatsn_nlya}(b). 
It is evident from this figure 
that the critical value of $\varepsilon$ for small noise intensity is almost the same as that of the oscillators without noise [cf. Figs.~\ref{adcatsn_lya}(a) 
and \ref{adcatsn_lya}(b)], indicating the robustness of
the synchronous state and the exponential relation (\ref{eqn:eps_c}) with 
the addition of small noise. Further the
\begin{figure}[!ht]
\centering\includegraphics[width=0.99\linewidth]{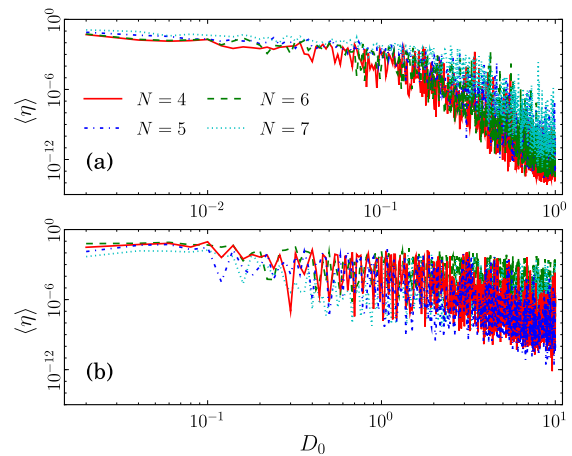}
\caption{(color online) Time averaged synchronization error $\langle \eta \rangle$ for different number of oscillators in the ring as a function of noise intensity $D_0$ for fixed value of $\varepsilon_c$ displaying power-law decay for (a) the R\"ossler oscillators in the range $D_0\in(0.01,1)$, and (b) the Lorenz oscillators in the range $D_0\in(0.01,10)$ with the second additional coupling at the second neighbor of the first coupling.} 
\label{adcatsn_ni}
\end{figure}
power law decay of the synchronization error as a function of the noise intensity [Figs.~\ref{adcatsn_ni}(a)-(b)] beyond certain threshold values of $D_0$
for the value of $\varepsilon_c$ for both the oscillators confirming the 
existence of noise-enhanced synchronization. Exactly similar figures can be 
obtained by fixing the coupling strength to be less than  $\varepsilon_c$ 
indicating the existence of noise-induced synchronization.

\subsection{Couplings at the $N$-th neighbors}

Let $N_c$ be the maximum number of oscillators in the response
array that are synchronized with the external drive for a given
$\varepsilon_c$ for $m=1$. Then, by
introducing a second coupling, $(m=2)$ in eq.~(\ref{eqn:p}), at
$N_c+1$ one can synchronize up to a maximum of $2N_c$ oscillators
in the response array with the external drive for the same coupling
strength $\varepsilon_c$. One can also introduce a third coupling
in the response array, $(m=3)$ in eq.~(\ref{eqn:p}), in the same
fashion in which the second coupling is introduced where one can
synchronize a maximum $3N_c$ oscillators in the response array. In
this way one can increase the number of oscillators (size) in
the response array that evolve in synchrony with the external drive
to $mN_c$ by introducing $m$ couplings at the oscillator
index $1$, $N_c+1$, $2N_c+1$, \ldots, $(m-1)N_c+1$,
respectively, with the same  critical coupling $\varepsilon_c$.

In the following we shall illustrate this result using the second and third couplings
at the oscillators with the indices $N_c+1$ and $2N_c+1$, respectively. 
It is known from Section~\ref{sec2} that the maximum number of oscillators
that can be synchronized with the drive with a single coupling is $N_c=5$. The CLEs
of the R\"ossler and Lorenz oscillators with the second coupling, $(m=2)$, at the oscillator in
the ring with the index $N_c+1=6$ and the third coupling, $(m=3)$, at the oscillator
index $2N_c+1=11$ along with the CLE of the $N_c$ oscillators in the ring with a single
\begin{figure}[!ht]
\centering\includegraphics[width=0.99\linewidth,clip]{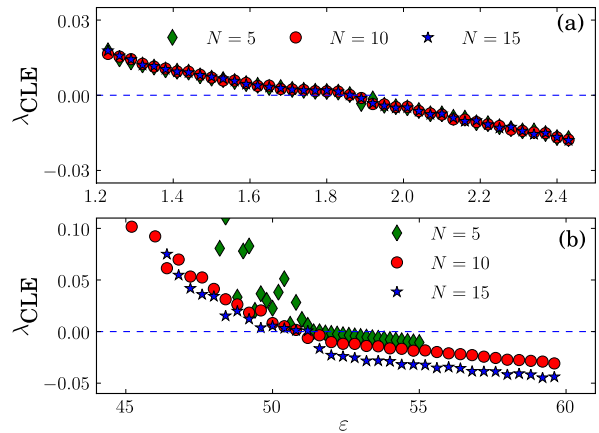}
\caption{(color online) Conditional Lyapunov exponents of the (a) R\"ossler and, (b) Lorenz system  for different values of the number of oscillators in the array as a function of the coupling strength with the second and the third additional couplings at the $N$-th and $2N$-th neighbor, respectively,
of the first coupling.} 
\label{adcatnn_lya}
\end{figure}
coupling are shown in Figs.~\ref{adcatnn_lya}(a) and \ref{adcatnn_lya}(b) as a function of $\varepsilon$. 
It clearly shows that the
$mN_c$ oscillators in the ring with couplings at $m=1,2$ and $3$ are synchronized exactly
at the same critical value of the coupling strength $\varepsilon_c$.
Hence, it is evident that
the number of synchronized oscillators can be increased beyond the size 
instability to any desired amount by introducing additional couplings at 
the $(mN_c+1)$-th oscillators for the same value of $\varepsilon_c$.

\begin{figure}[!ht]
\centering\includegraphics[width=0.99\linewidth,clip]{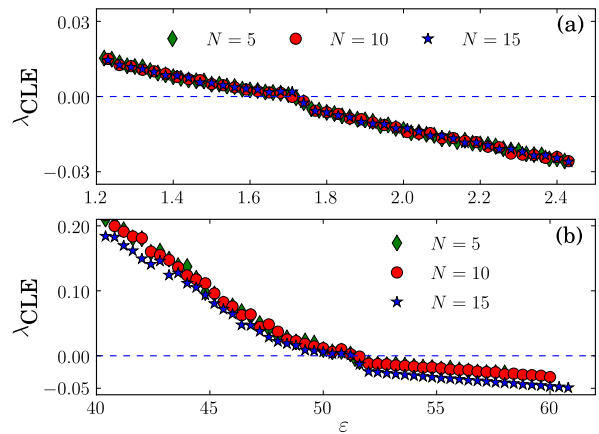}
\caption{(color online) Conditional Lyapunov exponents of the (a) R\"ossler and, (b) Lorenz system  
for different values of the number of oscillators in the array as a function 
of the coupling strength with second additional coupling at 
the $N$-th and $2N$-th neighbors, respectively,
of the first coupling in the presence of Gaussian white noise with
intensity $D_0=0.001$.} \label{adcatnn_nlya}
\end{figure}

\begin{figure}[!ht]
\centering\includegraphics[width=0.99\linewidth,clip]{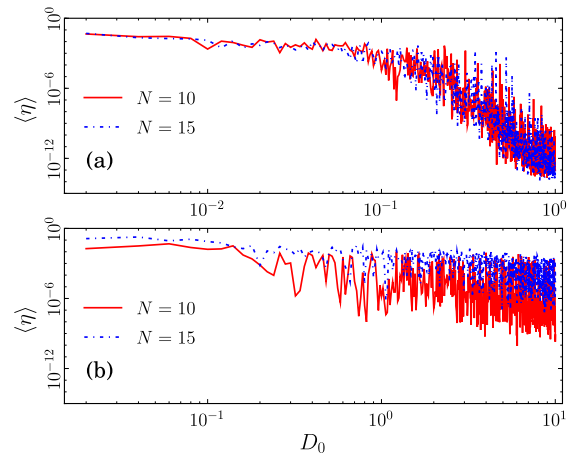}
\caption{(color online) Time averaged synchronization error $\langle \eta \rangle$ for different numbers of oscillators in the ring as a function of noise intensity $D_0$  for the fixed value of $\varepsilon_c$  displaying power-law decay for (a) the R\"ossler oscillators in the range $D_0\in(0.01,1)$, and (b) the
Lorenz oscillators in the range $D_0\in(0.01,10)$ with second additional coupling at the $N$-th and $2N$-th neighbor, respectively, of the first coupling.} 
\label{adcatnn_ni}
\end{figure}

We have also examined the effect of noise as in the previous cases. The CLEs of
the R\"ossler and Lorenz oscillators with the second and the third couplings
at the oscillators with the indices $N_c+1$ and $2N_c+1$ with the noise
intensity $D_0=0.001$ for fixed value of $\varepsilon_c$ are shown in
Figs.~\ref{adcatnn_nlya}(a) and~\ref{adcatnn_lya}(b), respectively. The synchronization
error also displays a power law decay as a function of the noise intensity $D_0$
as shown in Figs.~\ref{adcatnn_ni}(a) and \ref{adcatnn_ni}(b) 
beyond after certain threshold values of $D_0$
for both the coupled R\"ossler and Lorenz 
oscillators confirming the robustness of the synchronous states and the noise-enhanced synchronization. Further these systems of coupled oscillators also display noise-induced synchronization by exhibiting similar figures as a function of $D_0$ for $\varepsilon$ less than $\varepsilon_c$.

\subsection{Effect of additional couplings}

\begin{figure}[!ht]
\centering\includegraphics[width=0.99\linewidth,clip]{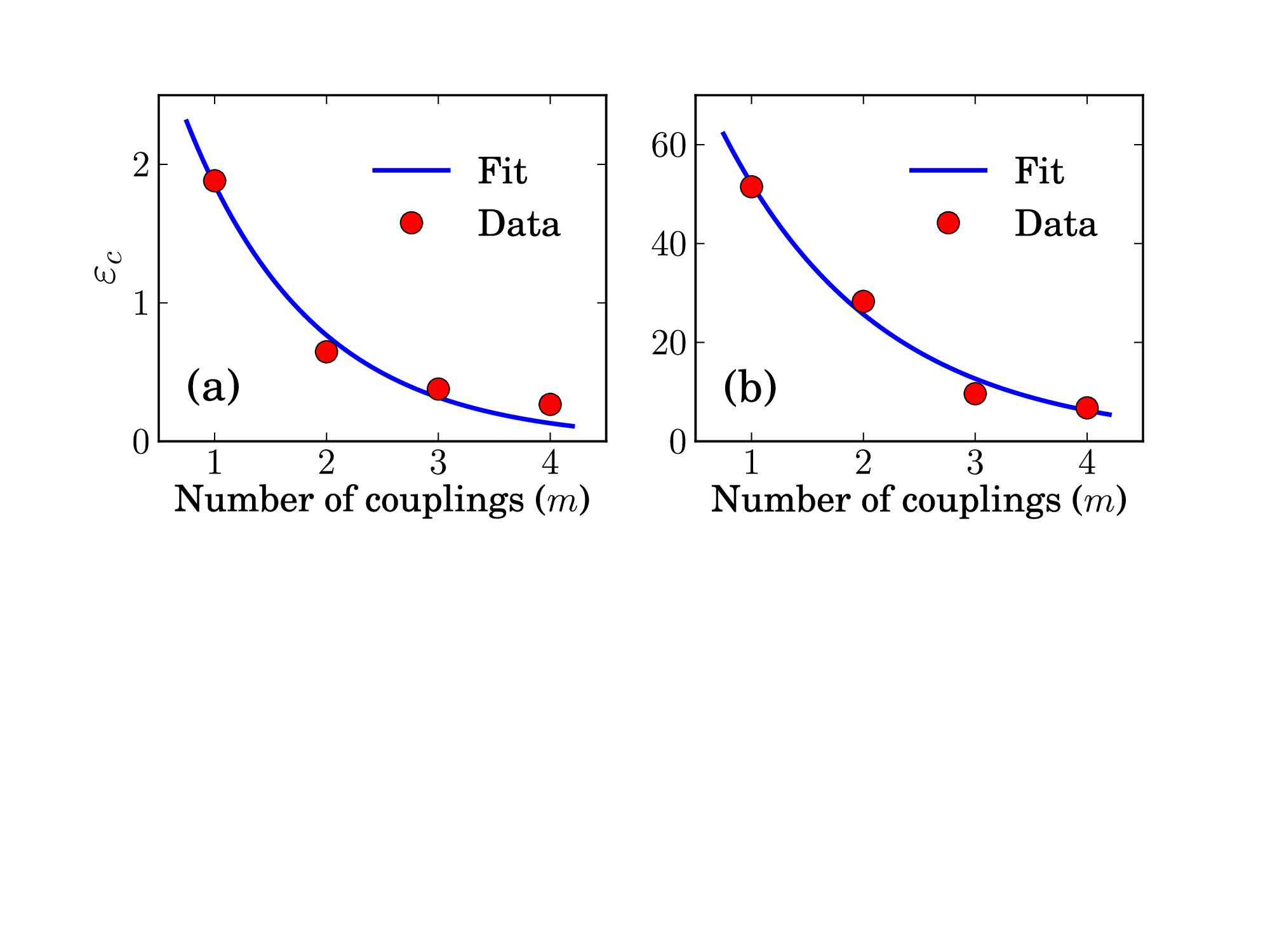}
\caption{(color online) Critical coupling strength as a function of the number of additional couplings between the drive and the response array for a fixed
number of oscillators in the array displaying an exponential decay of $\varepsilon_c$ as a function of number of additional couplings (a) R\"ossler and, 
(b) Lorenz system.} \label{ec_roslor}
\end{figure}

In addition to the effect of increasing the number of synchronized oscillators in the array
beyond the size instability limit by introducing additional couplings at $(mN_c+1)$-th oscillators
for $m=1,2,\cdots$ it reduces the required coupling strength exponentially to synchronize the 
fixed number of oscillators in the array. For instance, we have fixed the number of oscillators
in the array to be $N=N_c=5$ and increase the  number of couplings between the drive and
the array from $N=1$ to $N=5$-th oscillator and estimate the critical coupling 
strength  $\varepsilon_c$ required to synchronize the $N=N_c=5$ oscillators in the
array for each additional coupling. The estimated $\varepsilon_c$ is plotted as
a function of the number of additional couplings in the array in 
Figs.~\ref{ec_roslor} for both the R\"ossler  and the  Lorenz systems, which 
establishes an exponential decrease of the required coupling strength to synchronize
the fixed number of oscillators in the array by introducing additional couplings
between the drive and the array.
It may be noted that the critical coupling strength follows an exponential relation with the number of couplings as $\varepsilon_c \sim \varepsilon_0 \exp (-\gamma\, m)$, with $m$ being the number of couplings. The constants, in the case of R\"ossler equations, turn out to be $\varepsilon_0 = 4.4746$ and $\gamma = 0.8819$ while they are $\varepsilon_0 = 105.6375 $ and $\gamma = 0.7069$ for Lorenz equations.

\section{Summary and conclusion}
\label{sec4}

In this paper, we have studied chaos synchronization in arrays of 
diffusively coupled nonlinear oscillators  with a ring geometry driven externally by an identical oscillator. In particular, we have shown that the critical 
coupling strength required to synchronize the array with the external drive 
increases exponentially with a scaling exponent $\gamma \in(0.3,0.5)$ as a 
function of the number $N_c$ of the oscillators in the array. We have pointed 
out that as a consequence of the exponential relation, the maximum number of 
oscillators in the array that can evolve in synchrony with the external drive 
is limited. Further, we have shown that by introducing additional couplings 
between the external drive and the array at $(mN_c+1)$-th oscillators in the ring, 
one can proportionately increase the maximum number of oscillators that can 
evolve in synchrony with the drive. 
Further, we have obtained
the same exponential relation connecting the critical coupling strength
and the number of oscillators even after introducing the additional number of couplings.
Furthermore, we have found that
$\varepsilon_c$ establishes an exponential decay as a function of
the number of additional couplings between the drive and the response array
for a fixed number of oscillators in the array.
We have also examined the robustness of the results against 
noise of small intensity and found that the synchronization error
displays a power law decay as a function of the noise intensity
at $\varepsilon=\varepsilon_c$ 
indicating the existence of noise-enhanced and noise-induced synchronization 
for $\varepsilon<\varepsilon_c$ in all the cases. 

In addition, we have also obtained similar results as above in other
ubiquitous coupled nonlinear oscillators such as coupled MLC circuits, Chua's
circuits and Sprott oscillators as well, and also in a discrete system, namely
coupled logistic maps, thus confirming
the universality of the above results. One can also extend the same type of analysis
with other coupling configurations such as star-type, unidirectional, global,
 weighted coupling configurations, 2-d, 3-d lattices, etc.
We believe  that our results shed more light on controllability and synchronizability of
networks by introducing additional couplings at appropriate oscillators/nodes
with the less cost in terms of  the coupling strength.
Further investigations can be extended to networks, in particular to network with
community structure using pinning control and also with delay coupling.

\acknowledgments

DVS has been supported by Alexander von Humboldt Foundation. The work of PM is supported by Fast Track Scheme for Young Scientists of the Department of Science and Technology (DST), Government of India (Ref. No. SR/FTP/PS-79/2005). ML acknowledges the support from a DST sponsored IRHPA research project and DST Ramanna Program. JK has been supported by the projects ECONS (WGL) and
EU project No. 240763 PHOCUS(FP7-ICT-2009-C). 

%\bibliography{sync,chaos}

\begin{thebibliography}{33}
\expandafter\ifx\csname natexlab\endcsname\relax\def\natexlab#1{#1}\fi
\expandafter\ifx\csname bibnamefont\endcsname\relax
  \def\bibnamefont#1{#1}\fi
\expandafter\ifx\csname bibfnamefont\endcsname\relax
  \def\bibfnamefont#1{#1}\fi
\expandafter\ifx\csname citenamefont\endcsname\relax
  \def\citenamefont#1{#1}\fi
\expandafter\ifx\csname url\endcsname\relax
  \def\url#1{\texttt{#1}}\fi
\expandafter\ifx\csname urlprefix\endcsname\relax\def\urlprefix{URL }\fi
\providecommand{\bibinfo}[2]{#2}
\providecommand{\eprint}[2][]{\url{#2}}

\bibitem[{\citenamefont{Fujisaka and Yamada}(1983{\natexlab{a}})}]{fuji:83:01}
\bibinfo{author}{\bibfnamefont{H.}~\bibnamefont{Fujisaka}} \bibnamefont{and}
  \bibinfo{author}{\bibfnamefont{T.}~\bibnamefont{Yamada}},
  \bibinfo{journal}{Prog. Theor. Phys.} \textbf{\bibinfo{volume}{69}},
  \bibinfo{pages}{32} (\bibinfo{year}{1983}{\natexlab{a}}).

\bibitem[{\citenamefont{Fujisaka and Yamada}(1983{\natexlab{b}})}]{fuji:83:02}
\bibinfo{author}{\bibfnamefont{H.}~\bibnamefont{Fujisaka}} \bibnamefont{and}
  \bibinfo{author}{\bibfnamefont{T.}~\bibnamefont{Yamada}},
  \bibinfo{journal}{Prog. Theor. Phys.} \textbf{\bibinfo{volume}{70}},
  \bibinfo{pages}{1240} (\bibinfo{year}{1983}{\natexlab{b}}).

\bibitem[{\citenamefont{Pecora and Corroll}(1990)}]{pecora:90:01}
\bibinfo{author}{\bibfnamefont{L.~M.} \bibnamefont{Pecora}} \bibnamefont{and}
  \bibinfo{author}{\bibfnamefont{T.~L.} \bibnamefont{Carroll}},
  \bibinfo{journal}{Phys. Rev. Lett.} \textbf{\bibinfo{volume}{64}},
  \bibinfo{pages}{821} (\bibinfo{year}{1990}).

\bibitem[{\citenamefont{Pikovsky et~al.}(2001)\citenamefont{Pikovsky,
  Rosenblum, and Kurths}}]{Pikovsky_etal:book:01}
\bibinfo{author}{\bibfnamefont{A.}~\bibnamefont{Pikovsky}},
  \bibinfo{author}{\bibfnamefont{M.}~\bibnamefont{Rosenblum}},
  \bibnamefont{and} \bibinfo{author}{\bibfnamefont{J.}~\bibnamefont{Kurths}},
  \emph{\bibinfo{title}{Synchronization: A Universal Concept in Nonlinear
  Sciences}} (\bibinfo{publisher}{Cambridge University Press},
  \bibinfo{address}{Cambridge}, \bibinfo{year}{2001}).

\bibitem[{\citenamefont{Lakshmanan and Murali}(1996)}]{lakshman_murali:book:96}
\bibinfo{author}{\bibfnamefont{M.}~\bibnamefont{Lakshmanan}} \bibnamefont{and}
  \bibinfo{author}{\bibfnamefont{K.}~\bibnamefont{Murali}},
  \emph{\bibinfo{title}{Chaos in Nonlinear Oscillators: Controlling and
  Synchronization}} (\bibinfo{publisher}{World Scientific},
  \bibinfo{address}{Singapore}, \bibinfo{year}{1996}).

\bibitem[{\citenamefont{Bohr and Christensen}(1989)}]{bohr:89:01}
\bibinfo{author}{\bibfnamefont{T.}~\bibnamefont{Bohr}} \bibnamefont{and}
  \bibinfo{author}{\bibfnamefont{O.~B.} \bibnamefont{Christensen}},
  \bibinfo{journal}{Phys. Rev. Lett.} \textbf{\bibinfo{volume}{63}},
  \bibinfo{pages}{2161} (\bibinfo{year}{1989}).

\bibitem[{\citenamefont{Heagy et~al.}(1994)\citenamefont{Heagy, Carroll, and
  Pecora}}]{heagy:94:01}
\bibinfo{author}{\bibfnamefont{J.~F.} \bibnamefont{Heagy}},
  \bibinfo{author}{\bibfnamefont{T.~L.} \bibnamefont{Carroll}},
  \bibnamefont{and} \bibinfo{author}{\bibfnamefont{L.~M.}
  \bibnamefont{Pecora}}, \bibinfo{journal}{Phys. Rev. E}
  \textbf{\bibinfo{volume}{50}}, \bibinfo{pages}{1874} (\bibinfo{year}{1994}).

\bibitem[{\citenamefont{Kocarev and Parlitz}(1995)}]{kocarev:94:01}
\bibinfo{author}{\bibfnamefont{L.}~\bibnamefont{Kocarev}} \bibnamefont{and}
  \bibinfo{author}{\bibfnamefont{U.}~\bibnamefont{Parlitz}},
  \bibinfo{journal}{Phys. Rev. Lett.} \textbf{\bibinfo{volume}{74}},
  \bibinfo{pages}{5028} (\bibinfo{year}{1995}).

\bibitem[{\citenamefont{Francisco and Muruganandam}(2003)}]{gerson_anand:03-01}
\bibinfo{author}{\bibfnamefont{G.}~\bibnamefont{Francisco}} \bibnamefont{and}
  \bibinfo{author}{\bibfnamefont{P.}~\bibnamefont{Muruganandam}},
  \bibinfo{journal}{Phys. Rev. E} \textbf{\bibinfo{volume}{67}},
  \bibinfo{pages}{066204} (\bibinfo{year}{2003}).

\bibitem[{\citenamefont{Arenas et~al.}(2008)}]{arenas:08:01}
\bibinfo{author}{\bibfnamefont{A.} \bibnamefont{Arenas}},
\bibinfo{author}{\bibfnamefont{A.} \bibnamefont{Díaz-Guilera}},
\bibinfo{author}{\bibfnamefont{J.} \bibnamefont{Kurths}},
\bibinfo{author}{\bibfnamefont{Y.} \bibnamefont{Moreno}},
  \bibnamefont{and}
  \bibinfo{author}{\bibfnamefont{C.}~\bibnamefont{Zhou}},
  \bibinfo{journal}{Phys. Rep.} \textbf{\bibinfo{volume}{469}},
  \bibinfo{pages}{93} (\bibinfo{year}{2008}). 

\bibitem[{\citenamefont{Osipov}(1986)}]{osipov:07:01}
\bibinfo{author}{\bibfnamefont{G.~V.}~\bibnamefont{Osipov}},
\bibinfo{author}{\bibfnamefont{C.}~\bibnamefont{Zhou}}, \bibnamefont{and}
  \bibinfo{author}{\bibfnamefont{J.}~\bibnamefont{Kurths}},
  \emph{\bibinfo{title}{Synchronization in Oscillatory Networks
 }} (\bibinfo{publisher}{Springer},
  \bibinfo{address}{Berlin}, \bibinfo{year}{2007}).

\bibitem[{\citenamefont{Mat{\'{\i}}as et~al.}(1997)\citenamefont{Mat{\'{\i}}as,
  P{\'e}rez-Mu{\~n}uzuri, Lorenzo, Mari{\~n}o, and
  P{\'e}rez-Villar}}]{matias:97-01}
\bibinfo{author}{\bibfnamefont{M.~A.} \bibnamefont{Mat{\'{\i}}as}},
  \bibinfo{author}{\bibfnamefont{V.}~\bibnamefont{P{\'e}rez-Mu{\~n}uzuri}},
  \bibinfo{author}{\bibfnamefont{M.~N.} \bibnamefont{Lorenzo}},
  \bibinfo{author}{\bibfnamefont{I.~P.} \bibnamefont{Mari{\~n}o}},
  \bibnamefont{and}
  \bibinfo{author}{\bibfnamefont{V.}~\bibnamefont{P{\'e}rez-Villar}},
  \bibinfo{journal}{Phys. Rev. Lett.} \textbf{\bibinfo{volume}{78}},
  \bibinfo{pages}{219} (\bibinfo{year}{1997}).

\bibitem[{\citenamefont{Mat{\'{\i}}as and G{\"u\'e}mez}(1998)}]{matias:98-01}
\bibinfo{author}{\bibfnamefont{M.~A.} \bibnamefont{Mat{\'{\i}}as}}
  \bibnamefont{and}
  \bibinfo{author}{\bibfnamefont{J.}~\bibnamefont{G{\"u\'e}mez}},
  \bibinfo{journal}{Phys. Rev. Lett.} \textbf{\bibinfo{volume}{81}},
  \bibinfo{pages}{4124} (\bibinfo{year}{1998}).

\bibitem[{\citenamefont{Turing}(1952)}]{turing:52-01}
\bibinfo{author}{\bibfnamefont{A.~M.} \bibnamefont{Turing}},
  \bibinfo{journal}{Phil. Trans. R. Soc. London}
  \textbf{\bibinfo{volume}{B327}}, \bibinfo{pages}{37} (\bibinfo{year}{1952}).

\bibitem[{\citenamefont{Collins and Stewart}(1994)}]{collins:94-01}
\bibinfo{author}{\bibfnamefont{J.~J.} \bibnamefont{Collins}} \bibnamefont{and}
  \bibinfo{author}{\bibfnamefont{I.}~\bibnamefont{Stewart}},
  \bibinfo{journal}{Biol. Cybern} \textbf{\bibinfo{volume}{71}},
  \bibinfo{pages}{95} (\bibinfo{year}{1994}).

\bibitem[{\citenamefont{Reddy and Sen}(2004)}]{reddy_sen:04-01}
\bibinfo{author}{\bibfnamefont{R.} \bibnamefont{Dodla}},
\bibinfo{author}{\bibfnamefont{A.} \bibnamefont{Sen}} \bibnamefont{and}
  \bibinfo{author}{\bibfnamefont{G.~L.} \bibnamefont{Johnston}},
  \bibinfo{journal}{Phys. Rev.E} \textbf{\bibinfo{volume}{69}},
  \bibinfo{pages}{056217} (\bibinfo{year}{2004}).

\bibitem[{\citenamefont{Abrams and Strogatz}(2004)}]{abrams_strogatz:04-01}
\bibinfo{author}{\bibfnamefont{D.~M.} \bibnamefont{Abrams}} \bibnamefont{and}
  \bibinfo{author}{\bibfnamefont{S.~H.} \bibnamefont{Strogatz}},
  \bibinfo{journal}{Phys. Rev. Lett.} \textbf{\bibinfo{volume}{93}},
  \bibinfo{pages}{174102} (\bibinfo{year}{2004}).

\bibitem[{\citenamefont{Deng and Huang}(2002)}]{deng_huang:02-01}
\bibinfo{author}{\bibfnamefont{X.~L.} \bibnamefont{Deng}} \bibnamefont{and}
  \bibinfo{author}{\bibfnamefont{H.~B.} \bibnamefont{Huang}},
  \bibinfo{journal}{Phys. Rev. E.} \textbf{\bibinfo{volume}{65}},
  \bibinfo{pages}{055202(R)} (\bibinfo{year}{2002}).

\bibitem[{\citenamefont{Wang and Huang}(2005)}]{wang_huang:05-01}
\bibinfo{author}{\bibfnamefont{H.~J.} \bibnamefont{Wang}},
  \bibinfo{author}{\bibfnamefont{H.~B.} \bibnamefont{Huang}},\bibnamefont{and}
  \bibinfo{author}{\bibfnamefont{G.~X.} \bibnamefont{Qi}},
  \bibinfo{journal}{Phys. Rev. E.} \textbf{\bibinfo{volume}{71}},
  \bibinfo{pages}{015202(R)} (\bibinfo{year}{2005}).

\bibitem[{\citenamefont{Wang and Huang}(2008)}]{sande_soriano:08-01}
\bibinfo{author}{\bibfnamefont{G.} \bibnamefont{VanderSande}},
  \bibinfo{author}{\bibfnamefont{M.~C.} \bibnamefont{Soriano}},
 \bibinfo{author}{\bibfnamefont{I.} \bibnamefont{Fischer}},\bibnamefont{and}
  \bibinfo{author}{\bibfnamefont{C.~R.} \bibnamefont{Mirasso}},
  \bibinfo{journal}{Phys. Rev. E.} \textbf{\bibinfo{volume}{77}},
  \bibinfo{pages}{055202(R)} (\bibinfo{year}{2008}).

\bibitem[{\citenamefont{Jiang et~al.}(2003)\citenamefont{Jiang, Lozada-Cassou,
  and Vinet}}]{jiang_etal:03-01}
\bibinfo{author}{\bibfnamefont{Y.}~\bibnamefont{Jiang}},
  \bibinfo{author}{\bibfnamefont{M.}~\bibnamefont{Lozada-Cassou}},
  \bibnamefont{and} \bibinfo{author}{\bibfnamefont{A.}~\bibnamefont{Vinet}},
  \bibinfo{journal}{Phys. Rev. E.} \textbf{\bibinfo{volume}{68}},
  \bibinfo{pages}{065201(R)} (\bibinfo{year}{2003}).

\bibitem[{\citenamefont{Lorenzo et~al.}(1996)\citenamefont{Lorenzo, Mari{\~n}o,
  Perez-Mu{\~n}uzuri, Mat{\'{\i}}as, and P{\'e}rez-Villar}}]{lorenzo:96-01}
\bibinfo{author}{\bibfnamefont{M.~N.} \bibnamefont{Lorenzo}},
  \bibinfo{author}{\bibfnamefont{I.~P.} \bibnamefont{Mari{\~n}o}},
  \bibinfo{author}{\bibfnamefont{V.}~\bibnamefont{Perez-Mu{\~n}uzuri}},
  \bibinfo{author}{\bibfnamefont{M.~A.} \bibnamefont{Mat{\'{\i}}as}},
  \bibnamefont{and}
  \bibinfo{author}{\bibfnamefont{V.}~\bibnamefont{P{\'e}rez-Villar}},
  \bibinfo{journal}{Phys. Rev. E.} \textbf{\bibinfo{volume}{54}},
  \bibinfo{pages}{R3094} (\bibinfo{year}{1996}).

\bibitem[{\citenamefont{Harmov et~al.}(2006)\citenamefont{Hramov, Koronovskii,
  Kurovskaya, and Boccaletti}}]{harmov:06-01}
\bibinfo{author}{\bibfnamefont{A.~E.} \bibnamefont{Hramov}},
  \bibinfo{author}{\bibfnamefont{A.~A.} \bibnamefont{Koronovskii}},
  \bibinfo{author}{\bibfnamefont{M.~K.} \bibnamefont{Kurovskaya}},
  \bibnamefont{and}
  \bibinfo{author}{\bibfnamefont{S.}~\bibnamefont{Boccaletti}},
  \bibinfo{journal}{Phys. Rev. Let.} \textbf{\bibinfo{volume}{97}},
  \bibinfo{pages}{114101} (\bibinfo{year}{2006}).

\bibitem[{\citenamefont{Brandt et~al.}(2006)\citenamefont{Brandt, Dellen, and
  Wessel}}]{brandt:06-01}
\bibinfo{author}{\bibfnamefont{S.~F.} \bibnamefont{Brandt}},
  \bibinfo{author}{\bibfnamefont{B.~K.} \bibnamefont{Dellen}},
  \bibnamefont{and} \bibinfo{author}{\bibfnamefont{R.}~\bibnamefont{Wessel}},
  \bibinfo{journal}{Phys. Rev. Let.} \textbf{\bibinfo{volume}{96}},
  \bibinfo{pages}{034104} (\bibinfo{year}{2006}).

\bibitem[{\citenamefont{Boccaletti et~al.}(2002)\citenamefont{Boccaletti,
  Kurths, Osipov, Valladares, and Zhou}}]{boccaletti:02-01}
\bibinfo{author}{\bibfnamefont{S.}~\bibnamefont{Boccaletti}},
  \bibinfo{author}{\bibfnamefont{J.}~\bibnamefont{Kurths}},
  \bibinfo{author}{\bibfnamefont{G.}~\bibnamefont{Osipov}},
  \bibinfo{author}{\bibfnamefont{D.~L.} \bibnamefont{Valladares}},
  \bibnamefont{and} \bibinfo{author}{\bibfnamefont{C.~S.} \bibnamefont{Zhou}},
  \bibinfo{journal}{Phys. Rep.} \textbf{\bibinfo{volume}{366}},
  \bibinfo{pages}{1} (\bibinfo{year}{2002}).

\bibitem[{\citenamefont{Rosenblum and Pikovsky}(2004)}]{rosenblum:04-01}
\bibinfo{author}{\bibfnamefont{M.~G.} \bibnamefont{Rosenblum}}
  \bibnamefont{and} \bibinfo{author}{\bibfnamefont{A.~S.}
  \bibnamefont{Pikovsky}}, \bibinfo{journal}{Phys. Rev. Lett.}
  \textbf{\bibinfo{volume}{92}}, \bibinfo{pages}{114102}
  (\bibinfo{year}{2004}).

\bibitem[{\citenamefont{Soen et~al.}(1999)\citenamefont{Soen, Cohen, Lipson,
  and Braun}}]{soen:99-01}
\bibinfo{author}{\bibfnamefont{Y.}~\bibnamefont{Soen}},
  \bibinfo{author}{\bibfnamefont{N.}~\bibnamefont{Cohen}},
  \bibinfo{author}{\bibfnamefont{D.}~\bibnamefont{Lipson}}, \bibnamefont{and}
  \bibinfo{author}{\bibfnamefont{E.}~\bibnamefont{Braun}},
  \bibinfo{journal}{Phys. Rev. Let.} \textbf{\bibinfo{volume}{82}},
  \bibinfo{pages}{3556} (\bibinfo{year}{1999}).

\bibitem[{\citenamefont{Pecora and Corroll}(1991)}]{pecora2}
\bibinfo{author}{\bibfnamefont{L.~M.} \bibnamefont{Pecora}} \bibnamefont{and}
  \bibinfo{author}{\bibfnamefont{T.~L.} \bibnamefont{Carroll}},
  \bibinfo{journal}{Phys. Rev. A} \textbf{\bibinfo{volume}{44}},
  \bibinfo{pages}{2374} (\bibinfo{year}{1991}).

\bibitem[{\citenamefont{Pecora and Carroll}(1998)}]{pecora:98:01}
\bibinfo{author}{\bibfnamefont{L.~M.} \bibnamefont{Pecora}} \bibnamefont{and}
  \bibinfo{author}{\bibfnamefont{T.~L.} \bibnamefont{Carroll}},
  \bibinfo{journal}{Phys. Rev. Lett.} \textbf{\bibinfo{volume}{80}},
  \bibinfo{pages}{2109} (\bibinfo{year}{1998}).

\bibitem[{\citenamefont{Restrepo et~al.}(2004)\citenamefont{Restrepo, Ott, and
  Hunt}}]{restrepo:04:01}
\bibinfo{author}{\bibfnamefont{J.~G.} \bibnamefont{Restrepo}},
  \bibinfo{author}{\bibfnamefont{E.}~\bibnamefont{Ott}}, \bibnamefont{and}
  \bibinfo{author}{\bibfnamefont{B.~R.} \bibnamefont{Hunt}},
  \bibinfo{journal}{Phys. Rev. Lett.} \textbf{\bibinfo{volume}{93}},
  \bibinfo{pages}{114101} (\bibinfo{year}{2004}).

\bibitem[{\citenamefont{Muruganandam
  et~al.}(1999{\natexlab{a}})\citenamefont{Muruganandam, Murali, and
  Lakshmanan}}]{muruganandam:99:01}
\bibinfo{author}{\bibfnamefont{P.}~\bibnamefont{Muruganandam}},
  \bibinfo{author}{\bibfnamefont{K.}~\bibnamefont{Murali}}, \bibnamefont{and}
  \bibinfo{author}{\bibfnamefont{M.}~\bibnamefont{Lakshmanan}},
  \bibinfo{journal}{Int. J. Bifur. Chaos: Appl. Sci. Eng.}
  \textbf{\bibinfo{volume}{9}}, \bibinfo{pages}{805}
  (\bibinfo{year}{1999}{\natexlab{a}}).

\bibitem[{\citenamefont{Muruganandam}(1999)}]{muru:thesis}
\bibinfo{author}{\bibfnamefont{P.}~\bibnamefont{Muruganandam}}, Ph.D. thesis,
  \bibinfo{school}{Bharathidasan University} (\bibinfo{year}{1999}).

\bibitem[{\citenamefont{Palaniyandi et~al.}(2005)\citenamefont{Palaniyandi,
  Muruganandam, and Lakshmanan}}]{palani_muru_lak:05:01}
\bibinfo{author}{\bibfnamefont{P.}~\bibnamefont{Palaniyandi}},
  \bibinfo{author}{\bibfnamefont{P.}~\bibnamefont{Muruganandam}},
  \bibnamefont{and}
  \bibinfo{author}{\bibfnamefont{M.}~\bibnamefont{Lakshmanan}},
  \bibinfo{journal}{Phys. Rev. E.} \textbf{\bibinfo{volume}{72}},
  \bibinfo{pages}{037205} (\bibinfo{year}{2005}).

\bibitem[{\citenamefont{Rangarajan and Ding}(2002)}]{rangarajan:02:01}
\bibinfo{author}{\bibfnamefont{G.}~\bibnamefont{Rangarajan}} \bibnamefont{and}
  \bibinfo{author}{\bibfnamefont{M.}~\bibnamefont{Ding}},
  \bibinfo{journal}{Phys. Lett. A} \textbf{\bibinfo{volume}{296}},
  \bibinfo{pages}{204} (\bibinfo{year}{2002}).

\bibitem[{\citenamefont{Chen et~al.}(2003)\citenamefont{Chen, Rangarajan, and
  Ding}}]{chen:03:01}
\bibinfo{author}{\bibfnamefont{Y.}~\bibnamefont{Chen}},
  \bibinfo{author}{\bibfnamefont{G.}~\bibnamefont{Rangarajan}},
  \bibnamefont{and} \bibinfo{author}{\bibfnamefont{M.}~\bibnamefont{Ding}},
  \bibinfo{journal}{Phys. Rev. E} \textbf{\bibinfo{volume}{67}},
  \bibinfo{pages}{026209} (\bibinfo{year}{2003}).

\bibitem[{\citenamefont{Sparrow}(1982)}]{sparrow:1982}
\bibinfo{author}{\bibfnamefont{C.}~\bibnamefont{Sparrow}},
  \emph{\bibinfo{title}{The Lorenz Equations: Bifurcations, Chaos, and Strange
  Attractors}} (\bibinfo{publisher}{Springer-Verlag}, \bibinfo{address}{New
  York}, \bibinfo{year}{1982}).

\bibitem[{\citenamefont{Lorenz}(1963)}]{lorenz:1963}
\bibinfo{author}{\bibfnamefont{E.~N.} \bibnamefont{Lorenz}},
  \bibinfo{journal}{J. Atm. Phys.} \textbf{\bibinfo{volume}{357}},
  \bibinfo{pages}{130} (\bibinfo{year}{1963}).

\bibitem[{\citenamefont{Zhou et~al.}(2002)\citenamefont{Soen, Cohen, Lipson,
  and Braun}}]{zhou:02-01}
\bibinfo{author}{\bibfnamefont{C.}~\bibnamefont{Zhou}},
  \bibinfo{author}{\bibfnamefont{J.}~\bibnamefont{Kurths}},
  \bibinfo{author}{\bibfnamefont{I.~Z.}~\bibnamefont{Kiss}}, \bibnamefont{and}
  \bibinfo{author}{\bibfnamefont{J.~L.}~\bibnamefont{Hudson}},
  \bibinfo{journal}{Phys. Rev. Let.} \textbf{\bibinfo{volume}{89}},
  \bibinfo{pages}{014101} (\bibinfo{year}{2002}).

\bibitem[{\citenamefont{Zhou et~al.}(2002)\citenamefont{Soen, Cohen, Lipson,
  and Braun}}]{zhou:02-02}
\bibinfo{author}{\bibfnamefont{C.}~\bibnamefont{Zhou}},\bibnamefont{and}
  \bibinfo{author}{\bibfnamefont{J.}~\bibnamefont{Kurths}},
  \bibinfo{journal}{Phys. Rev. Let.} \textbf{\bibinfo{volume}{88}},
  \bibinfo{pages}{230602} (\bibinfo{year}{2002}).

\bibitem[{\citenamefont{Kaneko}(1986)}]{kaneko:86:01}
\bibinfo{author}{\bibfnamefont{K.}~\bibnamefont{Kaneko}},
  \emph{\bibinfo{title}{Collapse of Tori and Genesis of Chaos in Dissipative
  Systems}} (\bibinfo{publisher}{World Scientific},
  \bibinfo{address}{Singapore}, \bibinfo{year}{1986}).

\end{thebibliography}

\end{document}